\documentclass[twocolumn,pre,showpacs,floats,superscriptaddress]{revtex4}
\usepackage{epsfig}

\begin{document}

\title{Surface morphology coarsening in a nonlocal system}

\author{Mikhail Khenner}
\affiliation{
Department of Mathematics, Western Kentucky University, Bowling Green, KY 42101
}

\pacs{68.55.J,81.15.Aa,81.16.Dn}

\begin{abstract}
Direct comparison is made of the steady-sates and coarsening dynamics in a local system and its nonlocal generalization. The example system is the surface of a solid film in a strong electric field; the morphological evolution
of the surface is described,  in the long-wavelength approximation, by  the amplitude PDE for the film height function. 
It is shown
that the amplitude of the steady-state and the coarsening rate of the surface structure are very sensitive to the  radius of the long-range interaction, and that both quantities increase as the radius decreases.
\end{abstract}

\date{\today}
\maketitle

Nonlocal pattern forming systems with long-range interactions, described by a nonlinear partial integro-differential equations, are ubiquitous in science and engineering.
One can mention Rayleigh-Benard convection \cite{CE},  magnetoconvection \cite{DP}, surfactant-mediated interfacial flows \cite{BBP}, instabilities in plasma \cite{BK,BB}, 
flow of a film down an inclined plane in the electric field \cite{GC,DE}, vibrated layers of a granular material or viscoelastic fluid \cite{DL}, evaporation of liquid films \cite{SBB}, and reaction-diffusion systems \cite{ABV,ST}. 

In materials science, evolution of phases of a binary alloy is described by a Cahn-Hilliard (CH) equation for an order parameter; 
the impacts of long-range interactions on coarsening of the order parameter have been studied \cite{F,RS,GLWW,S}. Also, guided by their analysis of Asaro-Tiller-Grinfeld instability in heteroepitaxial solid films, 
Kassner and Misbah \cite{KM} proposed the ``generic" nonlocal amplitude equation, derived in the long-wave limit:
\begin{equation}
h_t = -\alpha h +h_{xx}+H\left[h_x\right]-2h_x^2+\beta\left(H\left[h_x\right]\right)^2,
\label{KMeqn}
\end{equation}
where $h(x)$ is the film height above the substrate, $\alpha,\; \beta$ are parameters, and
\begin{eqnarray}
H\left[f(x)\right] &=& \frac{1}{\pi}\mbox{p.v.}\int_{-\infty}^{\infty}\frac{f(y)}{x-y}dy \nonumber \\
&=& \frac{1}{\pi}\lim_{\epsilon\rightarrow 0^+}\int_{|y-x|>\epsilon} \frac{f(y)}{x-y}dy
\label{HT}
\end{eqnarray}
is the Hilbert transform on the real line (p.v. stands for Cauchy principal value).  Eq. (\ref{KMeqn}), being relevant to systems where the linear dispersion relation is quadratic,
is complementary to the Kuramoto-Sivashinsky equation.  Although these authors computed Eq. (\ref{KMeqn}) and determined that it describes a perpetual coarsening,
the account of their investigation is very brief (one figure displaying a qualitative behavior); in particular, they did not investigate how the non-local terms affect the coarsening exponents. Nor were the effects of the finite radius of the long-range interactions studied.
\begin{figure}[h]
\vspace{-1.6cm}
\centering
\includegraphics[width=3.0in]{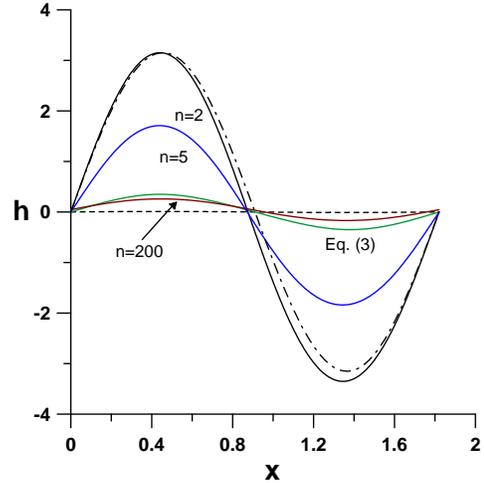}
\vspace{-1.2cm}
\caption{(Color online.) Steady-state surface profiles from the evolution of the one-wavelength, small-amplitude sinusoidal perturbation. This initial perturbation is shown by the dashed line.
$n=2, 5, 200$ curves are the steady-state profiles computed from Eq. (\ref{nonlocal-eq}). For comparison, the dash-dot line is the $a\sin{k_{max}x}$ curve, where $a$ is the height of the $n=2$ curve above $h=0$ level.
The steady-states were checked by computing the evolution towards equilibrium of the perturbed profiles.
}
\label{Fig_steady}
\end{figure}

In this short note, we directly compare coarsening in the local and nonlocal systems described by a long-wave evolution equations for the film height. These equations
stem from the consideration of the surface electromigration \cite{KD,SK,MB,DDF,BMOPL}, an intrinsically nonlocal effect. Assuming the constant electric field ${\mathbf E_0}$ 
parallel to the substrate and using the  
local aproximation for the field on the film surface, ${\mathbf E}={\mathbf E_0}\cos{\theta}$ (where $\theta$ is the surface orientation angle), the
\emph{local}, conserved evolution equation reads \cite{K}: 
\begin{eqnarray}
h_t &=& \frac{\partial}{\partial x}\left[-Bh_{xxx}+A\left\{-\frac{h_x^2}{2}+M'(0)\left(h_x-\frac{h_x^3}{2}\right)\right\}\right] \nonumber \\
& = & -Bh_{xxxx}+AM'(0)h_{xx}-Ah_xh_{xx} \nonumber \\ 
&-&\frac{3}{2}AM'(0)h_x^2h_{xx},
\label{local-eq}
\end{eqnarray}
where $B>0$ is the ``Mullins number" characterizing the strength of the natural surface diffusion, $A>0$ is the electric field strength parameter, and $M'(0)<0$ is the derivative of
the (anisotropic) adatom's diffusional mobility at the planar surface $h_x=0$. Under (dimensionless) Eq. (\ref{local-eq}) a random, short-wavelength initial deformation of the infinite planar 
surface perpetually coarsens - the size $L$ of the structures increases as a power law in time \cite{SK,K,KB}.
\begin{figure}[h]
\vspace{-1.6cm}
\centering
\includegraphics[width=3.0in]{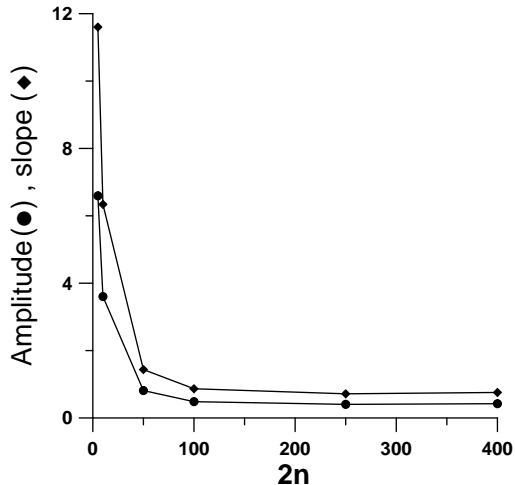}
\vspace{-1.2cm}
\caption{The amplitude and maximum slope of the steady-state profile vs. (twice) the interaction radius.
}
\label{Fig_ampl_slope}
\end{figure}

Guided by Ref. \cite{KM}, the \emph{nonlocal} generalization of this equation is:
\begin{eqnarray}
h_t &=& -Bh_{xxxx}+AM'(0)h_{xx}-AH_R\left[h_x\right]h_{xx} \nonumber \\
&-&\frac{3}{2}AM'(0)\left(H_R\left[h_x\right]\right)^2h_{xx},
\label{nonlocal-eq}
\end{eqnarray}
where 
\begin{equation}
H_R\left[f(x)\right] = \frac{1}{\pi}\int_{-R}^{R}\frac{f(y)}{x-y}dy,
\label{HT_finite}
\end{equation}
$R$ is the interaction radius. We compute Hilbert transform using fast method from Ref. \cite{BMB}.
The reason we focus on Eq. (\ref{nonlocal-eq}), rather than 
\begin{eqnarray}
h_t &=& \frac{\partial}{\partial x}\left[-Bh_{xxx}+A\left\{-\frac{H_R\left[h_x\right]^2}{2}\right.\right.\nonumber \\
&+&\left.\left.M'(0)\left(h_x-\frac{H_R\left[h_x\right]^3}{2}\right)\right\}\right]
\label{nonlocal-eq_other}
\end{eqnarray}
is because the latter equation describes the unlimited slope growth, and thus it does not allow to quantify the long-time effects of the nonlocal terms.
The unlimited slope growth is 
inconsistent with the fully nonlocal computations of the morphology evolution using the moving 
boundary problem \cite{SK} (where the electric field is computed in the bulk of the film using the Laplace equation for the electrical potential).

\begin{figure}[h]
\vspace{-1.6cm}
\centering
\includegraphics[width=3.0in]{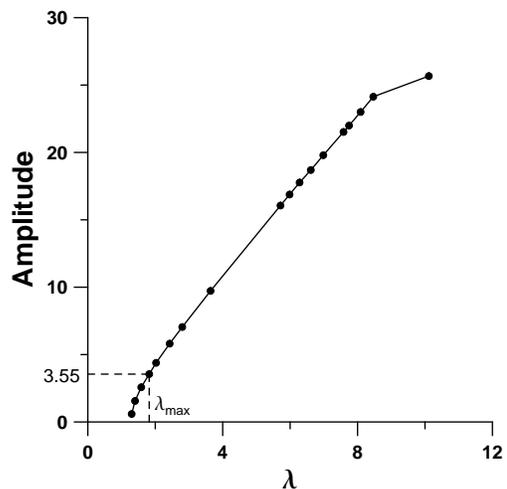}
\vspace{-1.1cm}
\caption{Amplitude of the steady-state surface profile vs. its imposed wavelength $\lambda$. 
$n=5$.
}
\label{ampl_vs_lambda}
\end{figure}

Fig. \ref{Fig_steady} shows the steady-state profiles from the initial condition $h(x,0)=0.01\sin{k_{max}x},\ 0\le x\le \lambda_{max}$, where $k_{max}=\sqrt{\frac{-AM'(0)}{2B}}$
is the most dangerous wavenumber from the linear stability analysis of Eqs. (\ref{local-eq}) and (\ref{nonlocal-eq}), and $\lambda_{max}$ is the corresponding wavelength.
The boundary conditions at $x=0, \lambda_{max}$ are periodic. 
$R$ in Fig. \ref{Fig_steady} is represented by one half of the number of the grid points, $n$, over which the integral in Eq. (\ref{HT_finite}) is computed (that is, to compute the transform
at $x_i$, the points $x_{-n},x_{-n+1},...,x_{i-1},x_i,x_{i+1},...,x_{n-1},x_{n}$ are used). $R=\infty$ (Eq. \ref{HT}) corresponds to all points involved, $n=200$.
It can be seen that when $R=\infty$ the steady-state amplitude is only slightly smaller than the one from the local Eq. (\ref{local-eq}), and the deviation from the local equation increases
as the radius 
\emph{decreases}.
To re-affirm these results, in Fig. \ref{Fig_ampl_slope}
we plot the steady-state amplitude and maximum slope vs. $R$; both quantities decrease fast initially and level off already at $n\sim 75$. And, from the 
monotonicity of the graph of the amplitude of the steady-state vs. the wavelength (Fig. \ref{ampl_vs_lambda}) one concludes that coarsening is 
uninterrupted in a nonlocal system \cite{PM1,PM2}. 

\begin{figure}[h]
\vspace{-1.6cm}
\centering
\includegraphics[width=3.0in]{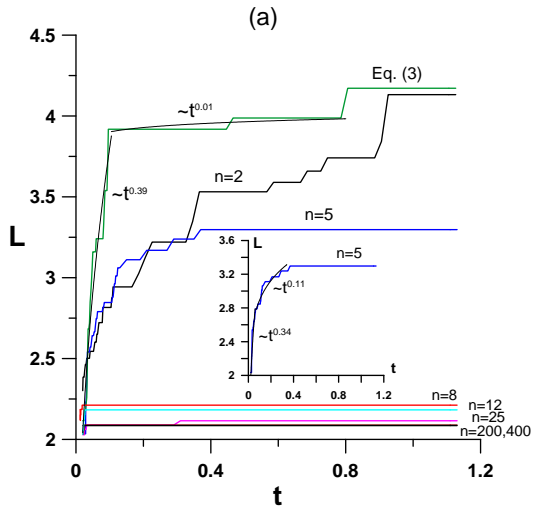}\\ \vspace{-2.5cm}\includegraphics[width=3.0in]{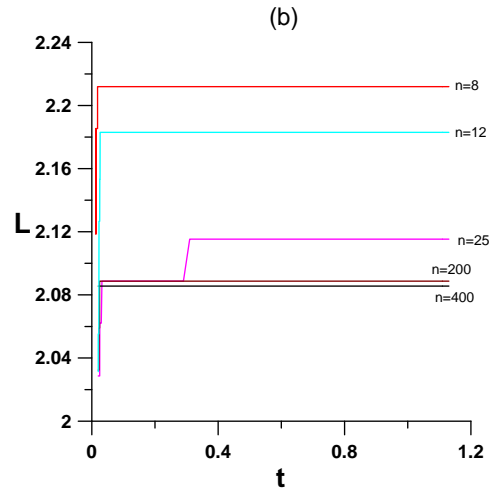}
\vspace{-1.2cm}
\caption{(Color online.) (a), (b):  Length scale of the coarsening surface structure vs. the time. (b) shows the zoom into (a).
The thin black lines in (a) are the power law fits to the data. 
}
\label{Fig_lengthscales}
\end{figure}

Next, we performed computations of coarsening using Eqs. (\ref{local-eq}) and (\ref{nonlocal-eq}). All such computations are done on the domain $0\le x \le 20\lambda_{max}$ with the periodic boundary conditions. The results are averaged over five runs with a different random initial condition,
and numerical convergence is checked on large grids. Here $R=\infty$ (Eq. \ref{HT}) corresponds to $n=400$. In Figures \ref{Fig_lengthscales}(a,b) $L$ is the horizontal scale of the surface  structure, e.g. 
the mean size of a hill at its base.  One can see that coarsening is very slow for large $R$ (or $n$) and speeds up when $R$ decreases, with the rate approaching one from the local equation as $R\rightarrow 0$.
The major speed-up occurs in a quenching fashion when  $n$  decreases from 8 to 5, and from there the speed-up is gradual. Also the coarsening rate is very weakly sensitive to $n$ for $n>\approx 10$ (Fig. \ref{Fig_lengthscales}(b)).

The coarsening exponent in the local model, Eq. (\ref{local-eq}), initially is of the order reported
in Ref. \cite{KD} for a similar local model (0.4 vs. 0.3), then it decreases sharply for the rest of the evolution. 
In the nonlocal
model, as $R$ increases, the initial regime is confined to the progressively shorter time intervals,
and then coarsening becomes logarithmically slow. The initial power law coarsening is matched to
the logarithmic law by a power law with a smaller exponent, as shown in the inset of Fig. \ref{Fig_lengthscales}(a).

\end{document}